\documentclass[]{aa}
\usepackage[longnamesfirst]{natbib}
\usepackage{graphicx}
\usepackage[a4paper]{hyperref}
\idline{000}{001}
\doi{2684}
\begin{document}
\def\teff{$T\rm_{eff }$}
\def\kms{${\mathrm {km s^{-1}}}$}
\headnote{Research Note}
\title{On the lithium content of the globular cluster M92
}
   \subtitle{}
\author{
P. \,Bonifacio \inst{1}
          }
\offprints{P. Bonifacio,\email{bonifaci@ts.astro.it}}
\institute{
Istituto Nazionale di Astrofisica --
Osservatorio Astronomico di Trieste, Via G.B.Tiepolo 11, 
I-34131 Trieste, Italy 
}
\authorrunning{Bonifacio}
\mail{P. Bonifacio}
\titlerunning{Lithium in M92}
\date{Received 14 May 2002  / Accepted  9 September 2002}
\abstract{
I use literature data and a new temperature
calibration to determine the Li abundances in
the globular cluster M 92.
Based on the same data, Boesgaard et al. have claimed
that there is a dispersion in Li abundances 
in excess of
observational errors. This result has been
brought as evidence for Li depletion in metal-poor dwarfs.
In the present  note I argue that there is no strong 
evidence for intrinsic dispersion in Li abundances, 
although a dispersion as large
as 0.18 dex is possible. 
The mean Li abundance, A(Li)=2.36, is in good 
agreement with recent results for field stars and
TO stars in the metal-poor globular cluster NGC 6397.
The simplest interpretation is that this constant value
represents the primordial Li abundance.
}
\maketitle{}           

\section{Introduction}

The present  note is motivated by the recent analysis
of the metal-poor globular cluster NGC 6397 by \citet{Bonifacio}.
We have confirmed that all the cluster stars share the same
Li abundance that is found exactly at the level
of the {\em Spite plateau} defined for field stars.
This  supports the notion that Li in metal-poor stars is
of primordial origin and good agreement is found with
the other primordial nuclei D and $\rm ^3He$. 
However, in this picture, the observations of 
\citet{boe98} of Li in the globular cluster
M 92 are troublesome.
After a careful analysis of the best available 
data, the above authors
conclude that there is a dispersion in Li abundances by a factor
of three, unexplained by observational errors.
The most likely explanation suggested for this dispersion
is differential Li depletion due to rotational mixing;
the stars would deplete more or less Li depending on their
initial angular momentum.
While this is a reasonable explanation, one is left to wonder
why such effect should be observable in M 92 but
not in NGC 6397, which has a very similar metallicity.
Moreover, recent models that predict Li depletion 
either through
rotational mixing \citep{pin01},
or diffusion
\citep{sal01},
or a combination of both plus composition gradient
 \citep{theado}, predict a mild depletion
of 0.1 -- 0.2 dex  
accompanied by  a very tiny dispersion. It  is
doubtful 
that such a tiny dispersion could be detected in M 92
with the presently available data.
The experience with NGC 6397 has shown that use of a photometric
colour such as $b-y$ or $B-V$ to estimate
effective temperatures is not well suited to the issue
of the scatter in Li abundances. The Li abundance is very
sensitive to errors in \teff , therefore 
quite small photometric errors in the colours, translate
into large errors on Li abundances.
In the case of NGC 6397, \citet{Bonifacio} 
took advantage of the fact that above the TO, where the
observed stars lie, there is a tight relation between V magnitude
and $b-y$; after correcting for the cluster reddening, this
may be calibrated against temperature.
The advantage is that an error of 0.05 mag in V translates into 
an error of only 20 -- 70 K in Teff.
A possible drawback is the vulnerability of the method to
binarity and \citet{boe98}  note that SH 18 may have a variable
radial velocity, indicating the presence of a companion.
In this  note I use the V -- \teff ~ calibration to
reanalyse the Li data in M 92. 
 
\begin{figure}[t!]
\resizebox{\hsize}{!}{\includegraphics[clip=true]{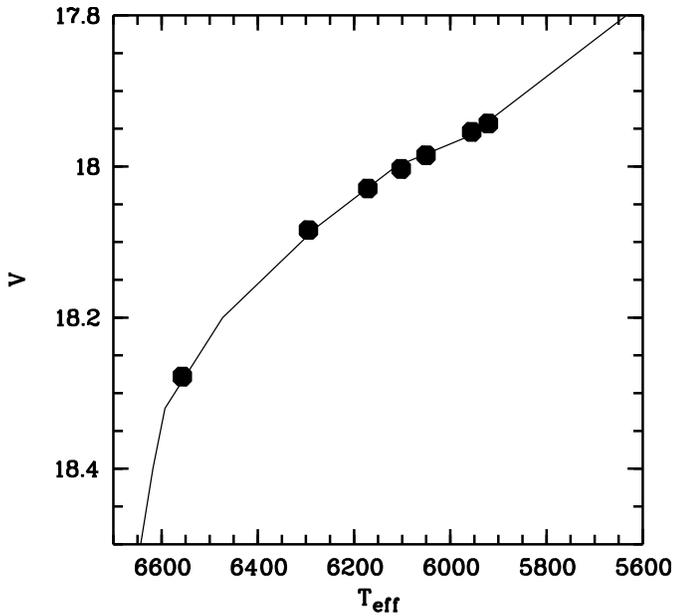}}
\caption{The adopted V -- \teff ~calibration: the solid line is 
the fiducial sequence for M92,  filled symbols correspond to the
star in Table \ref{liabun}}
\label{vteff}
\end{figure}

\begin{table*}
\caption{Equivalent widths and Li abundances for TO stars in M 92}
\label{liabun}
\begin{center}
\begin{tabular}{lrccrrrrrr}
\hline
\\
star  & EW &  $\sigma_{EW}$ & T$\rm _{eff}$ & A(Li) & $\sigma_{Li}$ & A(Li) &  A(Li)  & A(Li) & A(Li)\\
        & pm & pm             & K           &       &               &  TC            &  NLTE & B98 (K93) &  B98 (C83) \\
(1) & (2) & (3) & (4) & (5) & (6) & (7) & (8) & (9) (10) \\
\\
\hline
\\
SH 18  & 5.93   &  0.69 & 6050  & 2.53 & 0.07 & 2.55 & 2.57  & 2.57 & 2.45   \\
SH 21  & 2.50   &  0.62 & 5921  & 2.00 & 0.13 & 2.04 & 2.06  & 2.06 & 1.96   \\
SH 34  & 5.40   &  1.41 & 5956  & 2.38 & 0.17 & 2.42 & 2.43  & 2.36 & 2.22   \\
SH 46  & 2.87   &  0.74 & 6295  & 2.34 & 0.14 & 2.34 & 2.36  & 2.17 & 2.04   \\
SH 60  & 4.08   &  1.79 & 6171  & 2.51 & 0.23 & 2.52 & 2.54  & 2.42 & 2.30   \\
SH 80  & $<2.20$   &  1.10 & 6557  & $<2.39$ & 0.34  &  &    & $<2.23$ & $<2.16$   \\
SH 350 & 2.88   &  1.17 & 6102  & 2.20 & 0.25 & 2.22 & 2.23  & 2.13 & 2.01   \\
\\
\hline
\end{tabular}
\end{center}
\end{table*}

\begin{figure}[t!]
\resizebox{\hsize}{!}{\includegraphics[clip=true]{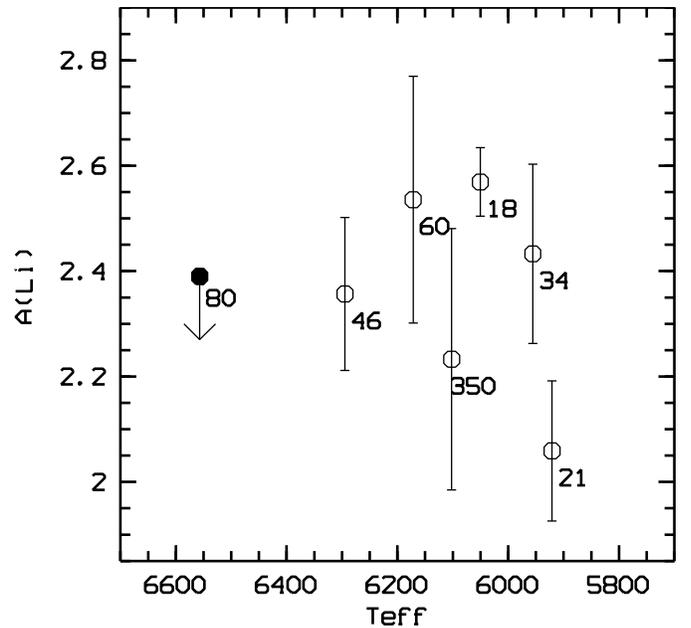}}
\caption{Li abundances, corrected for standard depletion and
NLTE effects, versus effective temperature for TO stars in M92.
The points are labelled with the star's SH number.
}
\label{li_teff}
\end{figure}

\section{Analysis}

I take the equivalent
widths of \cite{boe98}.
In order to use the $V - T_{eff}$
calibration of \citet{Bonifacio}, the cluster fiducial line
in the $V,(b-y)_0$ plane is required.
To assess the
scatter in Li abundances, a transformation 
of the the fiducial line of the cluster
of \citet{stetson} from the 
$(B-V),V$ plane to the $(b-y)_0,V$ plane will suffice.
For the cluster I adopt the reddening
$E(B-V)=0.025$ of \citet{Carretta}. 
A transformation from $(b-y)$ to $(B-V)$ 
has been published by \citet{Cousins},
however this cannot be readily inverted since it involves
also the metallicity-dependent index $m_0$.
To derive a transformation 
that applies to the TO stars of M92 I 
selected the stars used by \citet*{Bonifacio00} to determine
their intrinsic colour calibration, which have both Johnson and
Str\"omgren photometry  with $\rm -2.5\le [Fe/H]\le -1.5$.
The data for these stars is summarized in Table \ref{fieldata};
the photometric data has been slightly revised with
respect to \citet{Bonifacio00} according
to the current holdings of the General Catalog of Photometric data
\citep{GCPD}.
Both synthetic colours and Cousin's calibration suggest
that any  transformation between $(B-V)$ and $(b-y)$ 
should involve a metallicity-dependent term.
Since the Johnson system, unlike the Str\"omgren, 
does not have any specific metallicity dependent index and
the dependence of $U-B$ on metallicity in this range
of $B-V$ is as small as 
about 0.04 mag
per 0.5 dex in metallicity,
I prefer to use [Fe/H] directly as a parameter.
With as few as 11 calibrating points, it is not wise 
to fit anything more complex than a linear relation; however
the theoretical colours suggest that this should be adequate.
A $\chi^2$ fit of a linear
relation, taking into account errors in $B-V$,[Fe/H] (a 0.2 dex
error has been assumed for all the stars) and $b-y$
provides
$$ (b-y) = 0.74(B-V) -0.05[{\rm Fe/H}]
- 0.08 ;$$
the reduced $\chi^2$ is 0.88579 which denotes a good
fit and corresponds to a probability of 0.527, the root mean square
is 0.01 mag.
In the case of M92, [Fe/H]$\approx -2.0 $ so that the
transformation may be simplified to
$$ (b-y) \approx 0.74(B-V) + 0.00149 ;$$
This calibration is formally valid only for the colour and metallicity
range spanned by the calibrators, thus $0.357\le B-V \le 0.487$.
Three of the stars observed by \citet{boe98}
are slightly cooler than this range, however the largest extrapolation
required is for star \# 34 and it is of only 0.031 mag, and thus is 
not a serious problem for the issue addressed in this note.
\begin{table}
\caption{Data for the field stars used to derive the transformation}
\label{fieldata}
\begin{center}
\begin{tabular}{lrrrrr}
\hline
\\
star        &     $B-V$  & $\sigma_{B-V}$ & $b-y$ & $\sigma_{b-y}$ & [Fe/H]\\   
\hline
\\
G165-39       &  0.404  & 0.013  &0.309 & 0.001 & --2.05\\
G88-32        &   0.357 & 0.005  &0.309 & 0.006 & --2.36\\
G8-16         &   0.424 & 0.020  &0.322 & 0.006 & --1.59\\
HD 16031      &   0.441 & 0.003  &0.323 & 0.005 & --1.71\\
G13-35        &   0.432 & 0.007  &0.331 & 0.003 & --1.63\\
G59-24        &   0.423 & 0.008  &0.332 & 0.006 & --2.42\\
G11-44        &   0.434 & 0.011  &0.334 & 0.001 & --2.07\\
G114-26       &   0.481 & 0.007  &0.349 & 0.006 & --1.78\\
G37-26        &   0.459 & 0.009  &0.351 & 0.003 & --1.93\\
G60-48        &   0.487 & 0.004  &0.365 & 0.005 & --1.63\\
HD 34328      &   0.486 & 0.006  &0.365 & 0.007 & --1.61\\
\\
\hline
\end{tabular}
\end{center}
\end{table}

I use the above transformation to transpose
the fiducial TO of
\citet{stetson} 
to the $(b-y)_0, V$ plane. From this I obtain
the $V$ -- \teff calibration using the same
$b-y$ -- \teff used in \citet{Bonifacio} . 
With these effective temperatures I derive Li
abundances, as described in \citet{Bonifacio}:
I computed  a  model atmosphere 
for each temperature and log g = 3.75, as adopted
by \citet{King}.
The internal errors in effective temperature arising from
the  the photometric error in V are of the order of 10 to 50 K
and can be neglected in estimating the error on
A(Li)\footnote{I use the usual notation A(Li) = 12+log[N(Li)/N(H)]},
which is totally dominated by the errors on the equivalent widths.
Of course the external error is at least of the order of 100 K, 
and possibly larger, 
since I did not use Str\"omgren photometry;
however for the purpose of the
present discussion I shall not consider the external
error on effective temperatures.
Effective temperatures, Li abundances with their errors, as well as the 
equivalent   widths of \citet{boe98} as well as their Li abundances both
on the \citet{K93} and on the \citet{C83} scale,  are given
in Table \ref{liabun} .

There are 6 measures and one upper limit. Not unexpectedly the 
upper limit is for the hottest and faintest star of the sample,
therefore the one with the smallest equivalent width of the Li doublet
and among the smallest S/N ratios.
The mean A(Li) for the six measurements is 2.33 with 
a standard deviation of 0.20 . 
In order to assess if this
scatter is compatible with the observational error I use a Monte Carlo
simulation, as done by  \citet{Bonifacio}.
I generate 10000 samples of 7 ``observations'', each scattered
around a mean 2.3 with a Gaussian distribution of $\sigma$ equal
to the estimated error in A(Li) and compute the standard deviation
$s$ of these 7 ``observations''.
For the upper limit I consider an error of 0.34 dex, which is the
error corresponding to an error in equivalent width of 1.1 pm
for a star of \teff = 6557 K.
The mean $s$ of the 10000 samples is 0.20 with a standard deviation
of 0.07, therefore the simulations suggests that the observed scatter
is entirely due to the errors in the observations.  
I can also estimate the probability of having a non-detection of
Li in at least one star of the sample of 7. I define 
the minimum A(Li) detectable as 1.72, based on the data for 
SH 80, and find that out of 10000 samples, 400 satisfy this condition,
therefore the probability of having a non-detection is about 4\% .
This probability is small but non-negligible.
Of course since the errors are rather large (due to the faintness of 
the TO of M92) there could be an intrinsic scatter in Li 
abundances of the order of 
0.18 dex, which would go totally unnoticed.
One could suspect that the scatter found in the Monte Carlo
simulations is dominated by the inclusion 
of the rather large error associated with star SH 80, however this is
not the case. I performed a Monte Carlo simulation
with only 6 ``observations'' and the mean $s$ is
0.17 with a standard deviation of 0.06, therefore although
the scatter has undoubtedly decreased, at 1$\sigma$
this is consistent with the observed scatter.
One could further argue that 
the lack of detection of any extra scatter
is the result of considering also stars which have a very poor
Li abundance determination. If we, somewhat arbitrarily, select only the
stars for which the error on A(Li) is less than 0.2 dex we are left
with only 4 stars. The mean A(Li) is 2.31, not significantly changed,
and the dispersion is slightly larger: 0.22 dex.
A Monte Carlo simulation with 4 ``stars'' provides a mean 
standard deviation, over 10000 samples, 
of the four ``observations'' $<s> = 0.12$
with a standard deviation of 0.05 dex.
Now  an excess dispersion is detected, but at $2\sigma$ only.
At three $\sigma $ there is no evidence of extra dispersion.
One could claim that the existence of ``intrinsic dispersion''
is implied by the observation of SH 18 and SH 21, whose Li
abundance differs by 0.53 dex. 
In fact the two measures are compatible at 2.7 $\sigma$.
We may obtain an estimate of how likely this is
by looking at the Monte Carlo simulation: out of 10000 samples
in four realizations  the two ``stars'' differ
by more than this amount (in fact on one occasion
they differ by 0.65 dex).
This is a small probability; however whether one considers
this negligible or not must depend also on other information.

\begin{table}
\caption{Li curve of growth for \teff = 5951, log g = 3.75, [Fe/H]=--2.0}
\label{cog}
\begin{center}
\begin{tabular}{rrr}
\hline
\\
EW(pm) & log (EW)   A(Li)\\
\hline
2.09 &     0.320146 & 1.94\\
2.13 &     0.328380 & 1.95\\
2.64 &     0.421604 & 2.05\\
2.87 &     0.457882 & 2.09\\
3.61 &     0.557507 & 2.21\\
5.18 &     0.714330 & 2.39\\
8.00 &     0.903090 & 2.64\\
\hline
\end{tabular}
\end{center}
\end{table}

If we perform the corrections due to standard depletion and NLTE,
as described in \citet{Bonifacio}, the scatter in the data is diminished
by very little (0.19 dex rather than 0.20). 
However the mean A(Li) is now 2.36.
This value is
in good agreement with the Li abundance of field stars
\citep{b01} and of TO stars in NGC 6397 \citep{Bonifacio}.
The purpose of adopting the $V,T_{eff}$ calibration was to make
a direct comparison between M92 and NGC6397, however the 
dispersion is not decreased by the adoption of this temperature scale,
in fact the Li abundances of \citet{boe98} have a scatter of
0.20 dex on the \citet{K93} scale and 0.19 dex
on the \citet{C83} scale, essentially the same as found using
the $V,T_{eff}$ scale.
Performing a Monte Carlo simulation using 6 ``observations''
with the errors of \citet{boe98} I obtain a mean
$s$ of 0.12 dex with a standard deviation of 0.05 dex
for the \citet{K93} scale and 0.13 dex
 with a standard deviation of 0.05 dex for the \citet{C83} scale.
In this case extra dispersion is detected at 1$\sigma$ in both
cases, but in neither case at $2\sigma$.
The lower scatter is
due to the smaller errors adopted by
\citet{boe98} with respect to mine for several stars.
This is odd, since \citet{boe98} add  quadratically the 
errors due to equivalent widths and effective temperatures, whereas
I consider only the error due to the equivalent widths. Therefore,
my errors should be, if anything, smaller.
The only possible explanation is that there are some typographical
errors in table 2 of \citet{boe98}.
To check this I computed the curve of growth using a model
with \teff = 5951 K, which is adopted by \citet{boe98} for 
SH 46 on the \citet{K93} scale.
This is given in Table \ref{cog}.
From this table it can be seen that 
an error of 0.74 pm in equivalent width, around
an equivalent width of 2.87 pm induces an error of 0.14 dex
on the Li abundance. The  error given in Table 2 of
\citet{boe98} is 0.040 for the
\citet{K93} scale  and I think 
it is likely to be a typographical error.

\section{Conclusions}

From this reanalysis of existing data of M92 I conclude that there
is no strong support for an intrinsic dispersion in Li abundances;
however, a dispersion of the order of 0.18 dex would go unnoticed
with the present errors.
The mean Li abundance is in good agreement with that
found for field stars and for TO stars of NGC 6397, which has
a metallicity comparable to that of M92.
I believe that the simplest hypothesis is that Li is constant
in M92 and at the level of the {\em Spite plateau}.
However, the main purpose of this  note is to encourage observers
to obtain spectra with higher S/N ratio in order to check if this
is really the case. 
It is clear that statistical analysis and Monte Carlo simulations
cannot make up for inadequate observations.
The presently-available observations of
M 92 cannot prove nor disprove the existence of an ``intrinsic''
dispersion.
In particular, if SH 80 is really Li-depleted, 
it would be the first Li-depleted dwarf
so-far detected in a metal-poor globular
cluster and could testify in favour of Li-depletion
mechanisms in metal-poor dwarfs.
It is also obvious that an instrument like HIRES
cannot tackle the problem without a prohibitive
investment of telescope time. 
The new generation of medium resolution spectrographs
operating on 8m telescopes, like ESI at Keck
\citep{1998SPIE.3355...48E}, or
proposed, like AVES 
for VLT \citep{2000SPIE.4008..167P}, have the capability
to solve the problem.
For example, ESI, with an entrance slit of 0\farcs{3}
which projects to 1.95 pixels, provides 
a resolving power of 0.032 nm per resolution element,
which does not fully resolve the Li doublet, but is
enough to measure an accurate equivalent width.
There is no exposure time calculator for ESI, but
a V=10.55 star 
provides 56650 $e^{-}{\mathrm n \mathrm m} ^{-1} {\mathrm s}^{-1}$; scaling
this number for a V=18.278 star (SH 80) 
one obtains 46 $e^{-}{\mathrm n \mathrm m} ^{-1} {\mathrm s}^{-1}$,
therefore in one hour of exposure  the S/N $ \sim 73$ per resolution element.
Assuming one can reach a S/N $\sim  140$ adding 4 one hour exposures
the Cayrel formula \citep{cayrel88} provides an error on the equivalent
width of 0.24 pm, i.e. comparable to what was obtained
by \citet{Bonifacio} for the much brighter stars in 
NGC 6397. AVES has been proposed for the VLT and could not
obviously observe M92; however the design of AVES can easily
be modified for a different telescope and its performances
without adaptive optics would be similar to those of ESI
and better with adaptive optics.
Therefore it is from these medium resolution spectrographs that 
we must expect accurate abundances for TO stars in 
Galactic Globular Clusters in the near future.

\begin{acknowledgements}

This work was done while I was
at the Observatoire de Paris-Meudon as
a visitor.
R. Cayrel,  M. Spite and F. Spite
are warmly  thanked for many
helpful discussions. 
I am grateful to the referee, P.E. Nissen,
for pointing out the need for a 
colour--colour transformation valid for metal-poor stars.
This research was done with support from the 
italian MURST/MIUR COFIN2000 grant 
``Spettroscopia e studio dei processi fisici del Sole e delle 
stelle di tipo solare'' (P.I. G. Peres)

\end{acknowledgements}

\bibliographystyle{aa}

\begin{thebibliography}{}
\bibitem[{Boesgaard et al. (1998)}]{boe98} 
Boesgaard, A.\ M., Deliyannis, C.\ P., 
Stephens, A., \& King, J.\ R.\ 1998 ApJ, 493, 206 
\bibitem[Bonifacio
et al. (2000)]
{Bonifacio00} 
Bonifacio, P., Caffau, E., \& Molaro, P.\ 2000, \aaps, 145, 473 
\bibitem[{Bonifacio (2001)}]{b01}
Bonifacio, P. 2001, in The link between stars and
cosmology, Eds. M. Chavez  and  D. Mayya  ,  (Dordrecht: Kluwer) ,in press
\bibitem[{Bonifacio et al. (2002)}]{Bonifacio}
Bonifacio, P. et al., 2002 \aap, 390, 91 
\bibitem[Carney(1983)]{C83} Carney, B.~W.\ 1983, \aj, 88, 610 
\bibitem[Carretta,  et al. (2000)]{Carretta} 
Carretta, E., Gratton, R.~G., 
Clementini, G., \& Fusi Pecci, F.\ 2000, \apj, 533, 215
\bibitem[{Cayrel (1988)}]{cayrel88}
Cayrel R., 1988 in
The Impact of Very High S/N Spectroscopy
on Stellar Physics. Eds
G. Cayrel de Strobel and M. Spite (Dordrecht: Kluwer) 345
\bibitem[Cousins(1987)]{Cousins} Cousins, A.~W.~J.\ 1987, The 
Observatory, 107, 80
\bibitem[Epps \& Miller(1998)]{1998SPIE.3355...48E} Epps, H.~W.~\& Miller, 
J.~S.\ 1998, \procspie, 3355, 48. 
\bibitem[King(1993)]{K93} King, J.~R.\ 1993, \aj, 106, 1206 
\bibitem[King et al. (1998)]{King} 
King, J.~R., Stephens, A., Boesgaard, A.~M., \& 
Deliyannis, C.\ 1998, \aj, 115, 666
\bibitem[Mermillod et al. (1996)]{GCPD}
Mermillod J.C., Hauck B., Mermillod M., 1996,
General Catalog of Photometric Data,
http://obswww.unige.ch/gcpd/gcdp.html
\bibitem[Pallavicini et al.(2000)]{2000SPIE.4008..167P} Pallavicini, R.~et 
al.\ 2000, \procspie, 4008, 167. 
\bibitem[{Pinsonneault et al. (2002)}]{pin01}
Pinsonneault, M.~H., Steigman, G., 
Walker, T.~P., \& Narayanan, V.~K.\ 2002, \apj, 574, 398 
\bibitem[Salaris,  \& Weiss(2001)]{sal01} 
Salaris, M., \&  Weiss, A.\ 2001, A\&A, 376, 955 
\bibitem[Stetson \& Harris(1988)]{stetson} 
Stetson, P.~B.~\& Harris, W.~E.\ 1988, \aj, 96, 909
\bibitem[Th\' eado \& Vauclair (2001)]{theado}
Th\' eado, S.,  \& Vauclair, S. 2001 A\& A 375, 70


\end{thebibliography}

\end{document}